\documentclass{llncs}

\usepackage{latexsym}
\usepackage{amssymb}
\usepackage{amsfonts}

\def\01{\{0,1\}}
\newcommand{\eps}{\varepsilon}
\newcommand{\ket}[1]{|#1\rangle}
\newcommand{\bra}[1]{\langle#1|}
\newcommand{\ketbra}[2]{|#1\rangle\langle#2|}
\newcommand{\norm}[1]{\mbox{$\parallel{#1}\parallel$}}
\newcommand{\inp}[2]{\langle{#1}|{#2}\rangle} 
\newcommand{\Tr}{\mbox{\rm Tr}}
\newcommand{\weight}{\mbox{\rm weight}}
\newcommand{\rank}{\mbox{\rm rank}}

\begin{document}
\title{Lower Bounds on Matrix Rigidity via a Quantum Argument}
\author{Ronald de~Wolf\thanks{Supported by a Veni grant
from the Netherlands Organization for Scientific Research (NWO) and also partially supported by the European Commission
under the Integrated Projects RESQ, IST-2001-37559 and Qubit Applications (QAP) funded by the IST directorate as
Contract Number 015848.}
}
\institute{
        CWI, Kruislaan 413, 1098 SJ, Amsterdam, the Netherlands.\\
        {\tt rdewolf@cwi.nl}}
\maketitle

\begin{abstract}
The \emph{rigidity} of a matrix measures how many of its entries need
to be changed in order to reduce its rank to some value.
Good lower bounds on the rigidity of an explicit matrix would imply good
lower bounds for arithmetic circuits as well as for communication complexity.
Here we reprove the best known bounds on the rigidity of Hadamard 
matrices, due to Kashin and Razborov, using tools from \emph{quantum} computing.
Our proofs are somewhat simpler than earlier ones (at least for those
familiar with quantum) and give slightly better constants.
More importantly, they give a new approach to attack this 
longstanding open problem.\\[2mm]
\end{abstract}

\section{Introduction}

\subsection{Rigidity}

Suppose we have some $n\times n$ matrix $M$ whose rank we want to reduce.
The \emph{rigidity} of $M$ measures the minimal number $R$ of entries we
need to change in order to reduce its rank to $r$. Formally:
$$
R_M(r)=\min\{\weight(M-\widetilde{M}) \mid \rank(\widetilde{M})\leq r\},
$$
where ``\weight'' counts the number of non-zero entries.
Here the rank could be taken over any field of interest; 
in this paper we consider the complex field.
Roughly speaking, high rigidity means that $M$'s rank is robust against changes:
changes in few entries won't change the rank much.

Rigidity was defined by Valiant~\cite[Section~6]{valiant:rigidity} in the 1970s
with a view to proving circuit lower bounds.  In particular,
he showed that an explicit $n\times n$ matrix $M$ with $R_M(\eps n)\geq n^{1+\delta}$ 
for $\eps,\delta>0$ would imply that log-depth arithmetic circuits that
compute the linear map $M:\mathbb{R}^n\rightarrow\mathbb{R}^n$ need
superlinear circuit size. Clearly, $R_M(r)\geq n-r$ for every full-rank 
matrix, since reducing the rank by 1 requires changing at least 1 entry.
This bound is optimal for the identity matrix, but usually far from tight.
Valiant showed that most matrices have rigidity $(n-r)^2$, but 
finding an \emph{explicit} matrix with high rigidity has been open for decades.

A very natural and widely studied candidate for such a high-rigidity matrix
are the Hadamard matrices. A Hadamard matrix is an orthogonal $n\times n$ matrix $H$ 
with entries $+1$ and $-1$. Such matrices exist whenever $n$ is a power of 2,
but are conjectured to exist whenever $n$ is a multiple of 4. 
Suppose we have a matrix $\widetilde{H}$ differing from $H$ in
$R$ positions such that $\rank(\widetilde{H})\leq r$. The goal in proving
high rigidity is to lower bound $R$ in terms of $n$ and $r$.
Alon~\cite{alon:rigidity} proved $R=\Omega(n^2/r^2)$, which was reproved
by Lokam~\cite{lokam:rigidity} using spectral methods. 
Kashin and Razborov~\cite{kashin&razborov:rigidity} improved this to
$R=\Omega(n^2/r)$. This is currently the best known for Hadamard matrices.

In view of the difficulty in proving strong lower bounds on rigidity proper,
Lokam~\cite{lokam:rigidity} also introduced a relaxed notion of rigidity. 
This limits the \emph{size} of each change in entries to some parameter $\theta>0$. Formally
$$
R_M(r,\theta)=\min\{\weight(M-\widetilde{M}) \mid 
\rank(\widetilde{M})\leq r, \norm{M-\widetilde{M}}_\infty\leq\theta\},
$$
where $\norm{\cdot}_\infty$ measures the largest entry (in absolute value) of its argument.
For Hadamard matrices, Lokam proved the bound $R_H(r,\theta)=\Omega(n^2/\theta)$ 
if $\theta\leq n/r$ and $R_H(r,\theta)=\Omega(n^2/\theta^2)$ if $\theta>r/n$.
In particular, if entries can change at most by a constant then the rigidity is $\Omega(n^2)$.
For the case $\theta>r/n$, Kashin and Razborov~\cite{kashin&razborov:rigidity}
improved the bound to $R_H(r,\theta)=\Omega(n^3/r\theta^2)$.
Study of this relaxed notion of rigidity is further motivated by the fact that
stronger lower bounds would separate the communication complexity
versions of the classes PH and PSPACE~\cite{lokam:rigidity}.

Apart from Hadamard matrices, the rigidity of some other explicit matrices 
has been studied as well, sometimes giving slightly better bounds 
$R_M(r)=\Omega(n^2\log(n/r)/r)$, for instance for Discrete Fourier Transform
matrices~\cite{friedman:rigidity,sss:rigidity,lokam:vandermonde}.
Very recently, Lokam~\cite{lokam:quadratic} showed a near-optimal
rigidity bound $R_P(n/17)=\Omega(n^2)$ for the matrix $P$ whose entries
are the square roots of distinct primes, and proved an 
$\Omega(n^2/\log n)$ arithmetic circuit lower bound for the induced
linear map $P:\mathbb{R}^n\rightarrow\mathbb{R}^n$.
This matrix $P$, however, is ``less explicit'' than Hadamard matrices
and the rigidity bound has no consequences for communication complexity
because $P$ is not a Boolean matrix.
Moreover, the same circuit lower bound was already shown by 
Lickteig~\cite{lickteig:polynomen} 
(see also~\cite[Exercise~9.5]{bcs:algebraiccomplexity})
without the use of rigidity.

\subsection{Our contribution}

In this paper we give new proofs of the best known bounds on the rigidity of
Hadamard matrices, both the standard rigidity and the relaxed one:
\begin{itemize}
\item if $r\leq n/2$, then $\displaystyle R_H(r)\geq \frac{n^2}{4r}$
\item $\displaystyle R_H(r,\theta)\geq\frac{n^2(n-r)}{2\theta n+r(\theta^2+2\theta)}$
\end{itemize}
Our constant in the former bound is a bit better than the one 
of Kashin and Razborov~\cite{kashin&razborov:rigidity} (their proof gives $n^2/256r$),
while in the latter bound it is essentially the same.
However, we feel our proof technique is more interesting than our precise result. 
As detailed in Section~\ref{secidea},
the proof relies on interpreting an approximation $\widetilde{H}$
of the Hadamard matrix $H$ as a \emph{quantum communication system}, and then using
quantum information theory bounds from~\cite{nayak:qfa} to relate the rank 
of $\widetilde{H}$ to the quality of its approximation.%
\footnote{The connection between the Hadamard matrix and quantum 
communication was also exploited in the lower bound for the 
communication complexity of inner product by Cleve et~al.~\cite{cdnt:ip}.}
Actually our bounds hold for all so-called \emph{generalized} Hadamard matrices;
these are the orthogonal matrices where all entries have the same magnitude.
However, for definiteness we will state the results for Hadamard matrices only.

This paper fits in a recent but fast-growing line of research where results
about \emph{classical} objects are proved or reproved using \emph{quantum} 
computational techniques. Other examples of this are lower bounds for locally
decodable codes and private information 
retrieval~\cite{kerenidis&wolf:qldc,wehner&wolf:improvedldc}, classical proof systems 
for lattice problems derived from earlier quantum proof 
systems~\cite{aharonov&regev:latticeqnp,aharonov&regev:latticenpconp},
strong limitations on classical algorithms for local search~\cite{aaronson:localsearch} 
inspired by an earlier quantum computation proof, a proof that the 
complexity class PP is closed under intersection~\cite{aaronson:pp},
formula size lower bounds from quantum lower bounds~\cite{lls:advform},
and a new approach to proving lower bounds for classical circuit depth  
using quantum communication complexity~\cite{kerenidis:qcircuit}.
                                                                                
It should be noted that the use of quantum computing is not strictly
necessary for either of our results. The first is proved in two steps: 
(1) using the quantum approach we show that every $a\times b$ submatrix 
of $H$ has rank at least $ab/n$ and (2) using a non-quantum 
argument we show that an approximation $\widetilde{H}$ with small $R$ 
contains a large submatrix of $H$ and hence by (1) must have high rank.
The result of (1) was already proved by Lokam~\cite[Corollary~2.2]{lokam:rigidity}
using spectral analysis, so one may obtain the same result classically 
using Lokam's proof for (1) and our argument for (2).
Either way, we feel the proof is significantly simpler than that 
of Kashin and Razborov~\cite{kashin&razborov:rigidity}, 
who show that a random $a\times a$ submatrix of $H$ has 
rank $\Omega(a)$ with high probability.
In contrast, the quantum aspects of our proof for the bound on $R_H(r,\theta)$
cannot easily be replaced by a classical argument, but that proof 
is not significantly simpler than the one of Kashin and Razborov
(which uses the Hoffman-Wielandt inequality) and the constant is essentially the same.

Despite this, we feel our quantum approach has merit for two reasons. 
First, it unifies the two results, both of which are now proved from
the same quantum information theoretic idea.  And second, using
quantum computational tools gives a whole new perspective on 
the rigidity issue, and might just be the new approach we need to 
solve this longstanding open problem. Our hope is that these techniques 
not only reprove the best known bounds, but will also push them further.
In Section~\ref{secnonquantum} we discuss two non-quantum approaches to 
the rigidity issue that followed a first version of the present paper,
and point out ways in which our approach is stronger.

\section{Relation to Quantum Communication}\label{secidea}

Very briefly, an $r$-dimensional \emph{quantum state} is a unit vector of complex amplitudes,
written $\ket{\phi}=\sum_{i=1}^r\alpha_i\ket{i}\in\mathbb{C}^r$. 
Here $\ket{i}$ is the $r$-dimensional vector that has a 1 in its $i$th 
coordinate and 0s elsewhere. The inner product between $\ket{\phi}$
and $\ket{\psi}=\sum_{i=1}^r\beta_i\ket{i}$ is $\inp{\phi}{\psi}=\sum_i\alpha_i^*\beta_i$.
A \emph{measurement} is described by a set of positive semidefinite operators $\{E_i\}$ that sum to identity.
If this measurement is applied to some state $\ket{\phi}$, the 
probability of obtaining outcome $i$ is given by $\bra{\phi}E_i\ket{\phi}$. 
If $\{\ket{v_i}\}$ is an orthonormal basis, then a measurement in this basis
corresponds to the projectors $E_i=\ketbra{v_i}{v_i}$.
In this case the probability of outcome $i$ is $|\inp{v_i}{\phi}|^2$.
We refer to~\cite{nielsen&chuang:qc} for more details about quantum computing.
We use $\norm{E}$ to denote the operator norm (largest singular value) of a matrix $E$,
and $\Tr(E)$ for its trace (sum of diagonal entries).

Our proofs are instantiations of the following general idea, which relates 
(approximations of) the Hadamard matrix to quantum communication.
Let $H$ be an $n\times n$ Hadamard matrix.
Its rows, after normalization by a factor $1/\sqrt{n}$, form an orthonormal
set known as the \emph{Hadamard basis}. 
If Alice sends Bob the $n$-dimensional quantum state $\ket{H_i}$
corresponding to the normalized $i$th row of $H$, and Bob measures the 
received state in the Hadamard basis, then he learns $i$ with probability~1.

Now suppose that instead of $H$ we have some rank-$r$ $n\times n$ matrix
$\widetilde{H}$ that approximates $H$ in some way or other.
Then we can still use the quantum states $\ket{\widetilde{H}_i}$ 
corresponding to its normalized rows for quantum communication.
Alice now sends the state $\ket{\widetilde{H}_i}$.  Crucially,
she can do this by means of an $r$-dimensional quantum state.
Let $\ket{v_1},\ldots,\ket{v_r}$ be an orthonormal basis
for the row space of $\widetilde{H}$. In order to send 
$\ket{\widetilde{H}_i}=\sum_{j=1}^r \alpha_j\ket{v_j}$, Alice sends
$\sum_{j=1}^r \alpha_j\ket{j}$ and Bob applies the unitary map 
$\ket{j}\mapsto\ket{v_j}$ to obtain $\ket{\widetilde{H}_i}$. 
He measures this in the Hadamard basis, and now his probability 
of getting the correct outcome $i$ is 
$$
p_i=|\inp{H_i}{\widetilde{H}_i}|^2.
$$ 
The ``quality'' of these $p_i$'s correlates with the ``quality'' of $\widetilde{H}$:
the closer the $i$th row of $\widetilde{H}$ is to the $i$th row of $H$, 
the closer $p_i$ will be to 1.

Accordingly, Alice can communicate a random element $i\in[n]$ via an $r$-dimensional 
quantum system, with average success probability $p=\sum_{i=1}^n p_i/n$.
But now we can apply the following upper bound on the average success probability,
due to Nayak~\cite[Theorem~2.4.2]{nayak:qfa}:\footnote{NB: this is not the 
well-known and quite non-trivial random access code lower bound from the same paper, 
but a much simpler statement about average decoding probabilities.}
$$
p\leq \frac{r}{n}.
$$
Intuitively, the ``quality'' of the approximation $\widetilde{H}$,
as measured by the average success probability $p$, gives 
a lower bound on the required rank $r$ of $\widetilde{H}$.
In the next sections we instantiate this idea in two different ways to get our two bounds.

We end this section with a simple proof of Nayak's bound due to Oded Regev. 
In general, let $\ket{\phi_1},\ldots,\ket{\phi_n}$ be the $r$-dimensional
states encoding $1,\ldots,n$, respectively, and $E_1,\ldots,E_n$
be the measurement operators applied for decoding. Then, using that the
eigenvalues of $E_i$ are nonnegative reals and that the trace of a matrix 
is the sum of its eigenvalues:
$$
p_i=\bra{\phi_i}E_i\ket{\phi_i}\leq\norm{E_i}\leq \Tr(E_i)
$$
and
$$
\sum_{i=1}^n p_i\leq \sum_{i=1}^n \Tr(E_i)=\Tr\left(\sum_{i=1}^n E_i\right)=\Tr(I)=r.
$$

\section{Bound on $R_H(r)$}\label{secboundRHr}

The next theorem was proved by Lokam~\cite[Corollary~2.7]{lokam:rigidity}
using some spectral analysis. We reprove it here using a quantum argument.

\begin{theorem}[Lokam]
Every $a\times b$ submatrix $A$ of $H$ has rank $r\geq ab/n$.
\end{theorem}

\begin{proof}
Obtain rank-$r$ matrix $\widetilde{H}$ from $H$ by setting all entries outside of $A$ to 0.
Consider the $a$ quantum states $\ket{\widetilde{H}_i}$ corresponding to the nonempty rows;
they have normalization factor $1/\sqrt{b}$. For each such $i$, 
Bob's success probability is
$$
p_i=|\inp{H_i}{\widetilde{H}_i}|^2=\left|\frac{b}{\sqrt{bn}}\right|^2=\frac{b}{n}.
$$
But we're communicating one of $a$ possibilities
using $r$ dimensions, so Nayak's bound implies
$$
\frac{1}{n}\sum_{i=1}^n p_i=p\leq\frac{r}{a}.
$$
Combining both bounds gives the theorem.
\qed
\end{proof}

Surprisingly, Lokam's result allows us quite easily to derive 
Kashin and Razborov's~\cite{kashin&razborov:rigidity} bound on rigidity, 
which is significantly stronger than Lokam's (and Alon's).
We also obtain a slightly better constant than~\cite{kashin&razborov:rigidity}: 
their proof gives $1/256$ instead of our $1/4$.
This is the best bound known on the rigidity of Hadamard matrices.

\begin{theorem}
If $r\leq n/2$, then $R_H(r)\geq n^2/4r$.
\end{theorem}

\begin{proof}
Consider some rank-$r$ matrix $\widetilde{H}$ with at most $R=R_H(r)$ ``errors'' compared to $H$.
By averaging, there exists a set of $a=2r$ rows of $\widetilde{H}$ with at most
$aR/n$ errors. Now consider the submatrix $A$ of $\widetilde{H}$ consisting
of those $a$ rows and the $b\geq n-aR/n$ columns that have no errors in those $a$ rows. 
If $b=0$ then $R\geq n^2/2r$ and we are done, so we can assume $A$ is nonempty.
This $A$ is errorfree, hence a submatrix of $H$ itself, and the previous theorem implies
$$
r=\rank(\widetilde{H})\geq \rank(A)\geq \frac{ab}{n}\geq \frac{a(n-aR/n)}{n}.
$$
Rearranging gives the theorem.
\qed
\end{proof}

The condition $r\leq n/2$ is important here. If $H$ is symmetric then its 
eigenvalues are all $\pm\sqrt{n}$ (because $H^T H=nI$), so we can reduce the rank to $n/2$ by 
adding or subtracting the diagonal matrix $\sqrt{n}I$. 
This shows that $R_H(n/2)\leq n$.

\section{Bound on $R_H(r,\theta)$}

We now consider the case where the maximal change in entries of $H$ is bounded by $\theta$.

\begin{theorem}
$\displaystyle R_H(r,\theta)\geq\frac{n^2(n-r)}{2\theta n+r(\theta^2+2\theta)}$.
\end{theorem}

\begin{proof}
Consider some rank-$r$ matrix $\widetilde{H}$ with at most $R=R_H(r,\theta)$ errors, 
and $\norm{H-\widetilde{H}}_{\infty}\leq\theta$.
As before, define the quantum states corresponding to its rows:
$$
\ket{\widetilde{H}_i}=c_i\sum_{j=1}^n \widetilde{H}_{ij}\ket{j},
$$
where $c_i=1/\sqrt{\sum_j \widetilde{H}_{ij}^2}$ is a normalizing constant.
Note that $\sum_j \widetilde{H}_{ij}^2\leq (n-\Delta(H_i,\widetilde{H}_i))+\Delta(H_i,\widetilde{H}_i)(1+\theta)^2=n+\Delta(H_i,\widetilde{H}_i)(\theta^2+2\theta)$, 
where $\Delta(\cdot,\cdot)$ measures Hamming distance.
Bob's success probability $p_i$ is now
\begin{eqnarray*}
p_i & = & |\inp{H_i}{\widetilde{H}_i}|^2\\
    & \geq & \frac{c_i^2}{n}(n-\theta\Delta(H_i,\widetilde{H}_i))^2\\
    & \geq & c_i^2(n-2\theta\Delta(H_i,\widetilde{H_i}))\\
    & \geq & \frac{n-2\theta\Delta(H_i,\widetilde{H}_i)}{n+\Delta(H_i,\widetilde{H}_i)(\theta^2+2\theta)}.
\end{eqnarray*}
Since $p_i$ is a convex function of Hamming distance and the average $\Delta(H_i,\widetilde{H}_i)$ is $R/n$,
we also get a lower bound for the average success probability:
$$
p\geq 
\frac{n-2\theta R/n}{n+R(\theta^2+2\theta)/n}.
$$
Nayak's bound implies $p\leq r/n$. Rearranging gives the theorem.
\qed
\end{proof}

\noindent
For $\theta\geq n/r$ we obtain the second result of Kashin and 
Razborov~\cite{kashin&razborov:rigidity}:
$$
R_H(r,\theta)=\Omega(n^2(n-r)/r\theta^2).
$$
If $\theta\leq n/r$ we get an earlier result of Lokam~\cite{lokam:rigidity}:
$$
R_H(r,\theta)=\Omega(n(n-r)/\theta).
$$

\section{Non-Quantum Proofs}\label{secnonquantum}

Of course, quantum mechanical arguments like the above can always be stripped
of their quantum aspects by translating to the underlying linear algebra language,
thus giving a non-quantum proof.  In this section we discuss the relation between 
our proof and two recent non-quantum approaches to rigidity.  Both are 
significantly simpler than the Kashin-Razborov proofs~\cite{kashin&razborov:rigidity}.

\subsection{Midrijanis's proof}

After reading a first version of this paper, Midrijanis~\cite{midrijanis:rigidity} published
a very simple argument giving the same bound $R_H(r)\geq n^2/4r$ for the special class
of Hadamard matrices defined by $k$-fold tensor product of the basic $2\times 2$ matrix (so $n=2^k$)
$$
H_{2^k}=\left(\begin{array}{rr}1 & 1\\ 1 & -1\end{array}\right)^{\otimes k}.
$$
Let $r\leq n/2$ be a power of 2. This $H_{2^k}$ consists of $(n/2r)^2$ disjoint
copies of $\pm H_{2r}$ and each of those has full rank $2r$. Each of
those copies needs at least $r$ errors to reduce its rank to $r$, so we need
at least $(n/2r)^2r=n^2/4r$ errors to reduce the rank of $H_{2^k}$ to $r$.
Notice, however, that this approach only obtains bounds for the case where
$H$ is defined in the above manner.%
\footnote{It's not clear how new this proof is, see the comments at Lance Fortnow's weblog 
{\tt http://weblog.fortnow.com/2005/07/matrix-rigidity.html} }

\subsection{The referee's proof}\label{secnonquantumref}

An anonymous referee of an earlier version of this paper
suggested that the quantum aspects were essentially redundant and could
be replaced by the following spectral argument. 
Suppose for simplicity that the Hadamard matrix $H$ and its rank-$r$ approximation 
$\widetilde{H}$ have normalized rows, and as before let $\ket{H_i}$ and 
$\ket{\widetilde{H}_i}$ denote their rows.  
The Frobenius norm of a matrix $A$ is $\norm{A}_F=\sqrt{\sum_{i,j}A_{ij}^2}$.
We can factor $\widetilde{H}^*=DE$, where $D$ is an $n\times r$ matrix with orthonormal columns
and $E$ is an $r\times n$ matrix with $\norm{E}_F=\norm{\widetilde{H}}_F$.
Using the Cauchy-Schwarz inequality, we bound
\begin{eqnarray*}
\sum_{i=1}^n \inp{H_i}{\widetilde{H}_i} & =& \Tr(H\widetilde{H}^*)=\Tr(HDE)\\ 
                                        & \leq & \norm{HD}_{F}\cdot\norm{E}_{F}\\
                                        & = & \norm{D}_{F}\cdot\norm{E}_{F}\\
                                        & = & \sqrt{r}\cdot\norm{\widetilde{H}}_{F}.
\end{eqnarray*}
This approach is quite interesting.  It gives the same bounds when applied
to the two cases of this paper (where $\sum_i \inp{H_i}{\widetilde{H}_i}$ 
and $\norm{\widetilde{H}}_{F}$ are easy to bound),
with less effort than the Kashin-Razborov proofs~\cite{kashin&razborov:rigidity}.
However, it is not an unrolling of the quantum proof, since the latter
upper bounds the sum of \emph{squares} of the inner products:
$$
\sum_{i=1}^n |\inp{H_i}{\widetilde{H}_i}|^2 \leq r.
$$
An upper bound on the sum of squares implies a bound on the sum of inner products 
via the Cauchy-Schwarz inequality, but not vice versa. Thus, even though the 
two bounds yield the same results in the two cases treated here, 
the quantum approach is potentially stronger than the referee's.

\section{Discussion}

As mentioned in the introduction, this paper is the next in a recent line of 
papers about classical theorems with quantum proofs. So far, these results 
are somewhat \emph{ad hoc} and it is hard to see what unifies them other than
the use of some quantum mechanical apparatus.  A ``quantum method''
in analogy to the ``probabilistic method''~\cite{alon&spencer:probmethod}
is not yet in sight but would be a very intriguing possibility.
Using quantum methods as a mathematical proof tool shows the usefulness
of the study of quantum computers, quantum communication protocols, etc., 
irrespective of whether a large quantum computer will ever be built in the lab.
Using the methods introduced here to prove stronger rigidity lower bounds
would enhance this further.

Most lower bounds proofs for the rigidity of a matrix $M$ in the literature (including ours) 
work in two steps: (1) show that all or most submatrices of $M$ have 
fairly large rank, and (2) show that if the number of errors $R$ is small, 
there is some (or many) big submatrix of $\widetilde{M}$ that is uncorrupted. 
Such an uncorrupted submatrix of $\widetilde{M}$ is a submatrix of $M$ 
and hence by (1) will have fairly large rank.
As Lokam~\cite{lokam:vandermonde} observes, this approach will not 
yield much stronger bounds on rigidity than we already have: it is easy 
to show that a random set of $R=O(\frac{\max(a,b)n^2}{ab}\log(n/\max(a,b)))$
positions hits every $a\times b$ submatrix of an $n\times n$ matrix.
Lokam's~\cite{lokam:quadratic} recent $\Omega(n^2)$ rigidity bound for 
a matrix consisting of the roots of distinct primes indeed does something
quite different, but unfortunately this technique will not work for matrices
over $\{+1,-1\}$ like Hadamard matrices.

To end this paper, let me describe two vague directions for improvements.
First, the approach mentioned above finds a submatrix of rank at least $r$ in $\widetilde{M}$ 
and concludes from this that $\widetilde{M}$ has rank at least $r$.
However, the approach usually shows that \emph{most} submatrices of 
$\widetilde{M}$ of a certain size have rank at least $r$.
If we can somehow piece these lower bounds for many submatrices 
together, we could get a higher rank bound for the matrix $\widetilde{M}$
as a whole and hence obtain stronger lower bounds on rigidity.

A second idea that might give a stronger lower bound for $R_H(r)$ is the following.
We used the result that every $a\times b$ submatrix of $H$ has rank at least $ab/n$.
This bound is tight for some submatrices but too weak for others. We conjecture
(or rather, hope) that submatrices for which this bound \emph{is} more or less tight,
are very ``redundant'' in the sense that each or most of its rows are spanned
by many sets of rows of the submatrix. Such a submatrix can tolerate
a number of errors without losing much of its rank, so then we don't need 
to find an uncorrupted submatrix of $\widetilde{H}$ (as in the current proof), 
but could settle for a submatrix with little corruption.

\subsection*{Acknowledgments}
Thanks to Satya Lokam for sending me a draft of~\cite{lokam:quadratic}
and for some helpful explanations, to Oded Regev and Gatis Midrijanis
for useful discussions, to Falk Unger for proofreading, and to the 
anonymous STACS'06 referee for the proof in Section~\ref{secnonquantumref}.

\bibliographystyle{splncs}

\end{document}